\documentclass[letterpaper,english,pra,aps,reprint,longbibliography,footinbib,amsfonts,amsmath,amssymb,footinbib,floatfix]{revtex4-1}
\usepackage[T1]{fontenc}
\usepackage[latin9]{inputenc}
\usepackage{color}
\usepackage{amsmath}
\usepackage{amssymb}
\usepackage{graphicx}
\usepackage{esint}

\makeatletter

\pdfpageheight\paperheight
\pdfpagewidth\paperwidth

\usepackage{babel}
\usepackage{ulem}
\usepackage{filemod}

\usepackage{babel}

\usepackage{babel}

\makeatother

\usepackage{babel}
\begin{document}

\title{Simulating Dirac models with ultracold atoms in optical lattices}

\author{Jean Claude Garreau}

\affiliation{Universit\'e de Lille, CNRS, UMR 8523 - PhLAM - Laboratoire de Physique
des Lasers Atomes et Mol\'ecules, F-59000 Lille, France}
\homepage{www.phlam.univ-lille1.fr/atfr/cq}

\author{V\'eronique Zehnl\'e}

\affiliation{Universit\'e de Lille, CNRS, UMR 8523 - PhLAM - Laboratoire de Physique
des Lasers Atomes et Mol\'ecules, F-59000 Lille, France}
\homepage{www.phlam.univ-lille1.fr/atfr/cq}

\date{\today}
\begin{abstract}
We present a general model allowing ``quantum simulation'' of one-dimensional
Dirac models with 2- and 4-component spinors using ultracold atoms
in driven 1D tilted optical latices. The resulting Dirac physics is
illustrated by one of its well-known manifestations, \emph{Zitterbewegung}.
This general model can be extended and applied with great flexibility
to more complex situations. 
\end{abstract}

\keywords{Dirac equation, ultracold atoms, optical lattices, quantum simulators}
\maketitle

\section{\label{sec:Intro}Introduction}

The Dirac theory of the electron (with its quantum-electrodynamical
corrections) is the most complete, precise, and experimentally well-tested
theory in physics. It combines quantum mechanics and relativistic
covariance in a general frame, automatically including the spin degree
of freedom, and predicting the existence of the positron. However,
in atomic physics, and \emph{a fortiori} in cold-atom physics, Dirac
theory has played a relatively restricted role, because, experimentally,
its domain of application ($v\sim c$) is not often attained (except
for inner-shell electrons of heavy atoms) and, theoretically, many
of its important results (e.g. fine structure) can be calculated with
a good precision in the simpler frame of Pauli theory (that is, Schr\"odinger
equation plus spin 1/2), at least for light atoms.

Recently, \emph{quantum simulation}~\citep{Georgescu:QuantumSimulation:RMP14}
became a mainstream in ultracold-atom physics~\citep{Bloch:QuantumSimulationsUltracoldGases:NP14}.
The basic idea, inspired by early Feynman insights~\citep{Feynman:SimulatingPhysics:IJTP82},
is to generate the physical behavior corresponding to some model,
e.g. condensed matter's Hubbard Hamiltonians, by ``artificially''
creating a corresponding Hamiltonian in more controlled conditions,
e.g. ultracold atoms in optical lattices~\citep{Bloch:ManyBodyUltracold:RMP08}.
This ``Hamiltonian engineering'' has been pushed quite far, with
the introduction of artificial gauge fields~\citep{Dalibard:ArtificialGaugePotentials:RMP11},
spin-orbit couplings and Dirac equation simulations~\cite{Gerritsma:QuantumSimulationDirac:N10,Witthaut:EffectiveDiracDynamicsBichrOptLatt:PRA11,Salger:KleinTunnelingBECOptLatt:PRL11,Galitski:SpinOrbitCouplingQuantumGases:N13},
quantum magnetism of neutral atoms~\citep{Lin:SyntheticMagneticFieldsForUltracold:N09,Struck:TunableGaugePotentialDrivenLattices:PRL12},
and the physics of disordered systems~\citep{Chabe:Anderson:PRL08,Billy:AndersonBEC1D:N08,Roati:AubryAndreBEC1D:N08,Kondov:ThreeDimensionalAnderson:S11,Manai:Anderson2DKR:PRL15}.

Quantum simulation of Dirac physics has benefit of a large interest
in recent years. This can be done in condensed matter systems by taking
advantage of the flexible concept of quasi-particles, where in particular
the Weyl semimetal~\cite{Wang:QuantumTransportDiracAndWeylReview:17}
is a pertinent concept, and recently the existence of ``type-II''
Weyl particles (that is a Weyl particle breaking Lorentz isotropy)~\cite{Soluyanov:TypeIIWeyl:N17}
has been suggested. Dirac quantum simulators using ion traps have
also been proposed~\cite{Lamata:RelativisticQuantumMechanicsTrappedIons:NJP11}.
Another popular way of quantum-simulating Dirac physics is by using
ultracold atoms in optical lattices, pioneered by Gerritsma \emph{et
al}.~\cite{Gerritsma:QuantumSimulationDirac:N10,Gerritsma:QuantumSimulationKleinParadox:PRL11},
who studied the phenomenon of Klein tunneling, also studied in refs.~\cite{Salger:KleinTunnelingBECOptLatt:PRL11,Witthaut:EffectiveDiracDynamicsBichrOptLatt:PRA11,Suchet:AnalogSimulationWeylParticles:EPL16}.
Without trying to be exhaustive, a wealth of interesting related phenomena
can also be studied: topological insulators, Dirac cones, spin-orbit
coupling, and even cyclotron dynamics~\cite{Mazza:OpticalLatticeBasedQuantumSimulator:NJP12,Kolovsky:WannierStarkStatesAndBlochOsc:PRA13,Tarruell:MergingDiracPtsHoneycombLattice:N12,Lopez-Gonzalez:EffectiveDiracEquationOptLatt:PRA14,JimenezGarcia:TunableSpinOrbitCouplingDrivingUltracold:PRL15,Zhang:RelativisticQuantumEffectsOfDiracUltracold:FPH12,Kolovsky:SimulatingCyclotronBlochDyn:FPH12}.

\begin{figure}
\begin{centering}
\includegraphics[width=0.9\columnwidth]{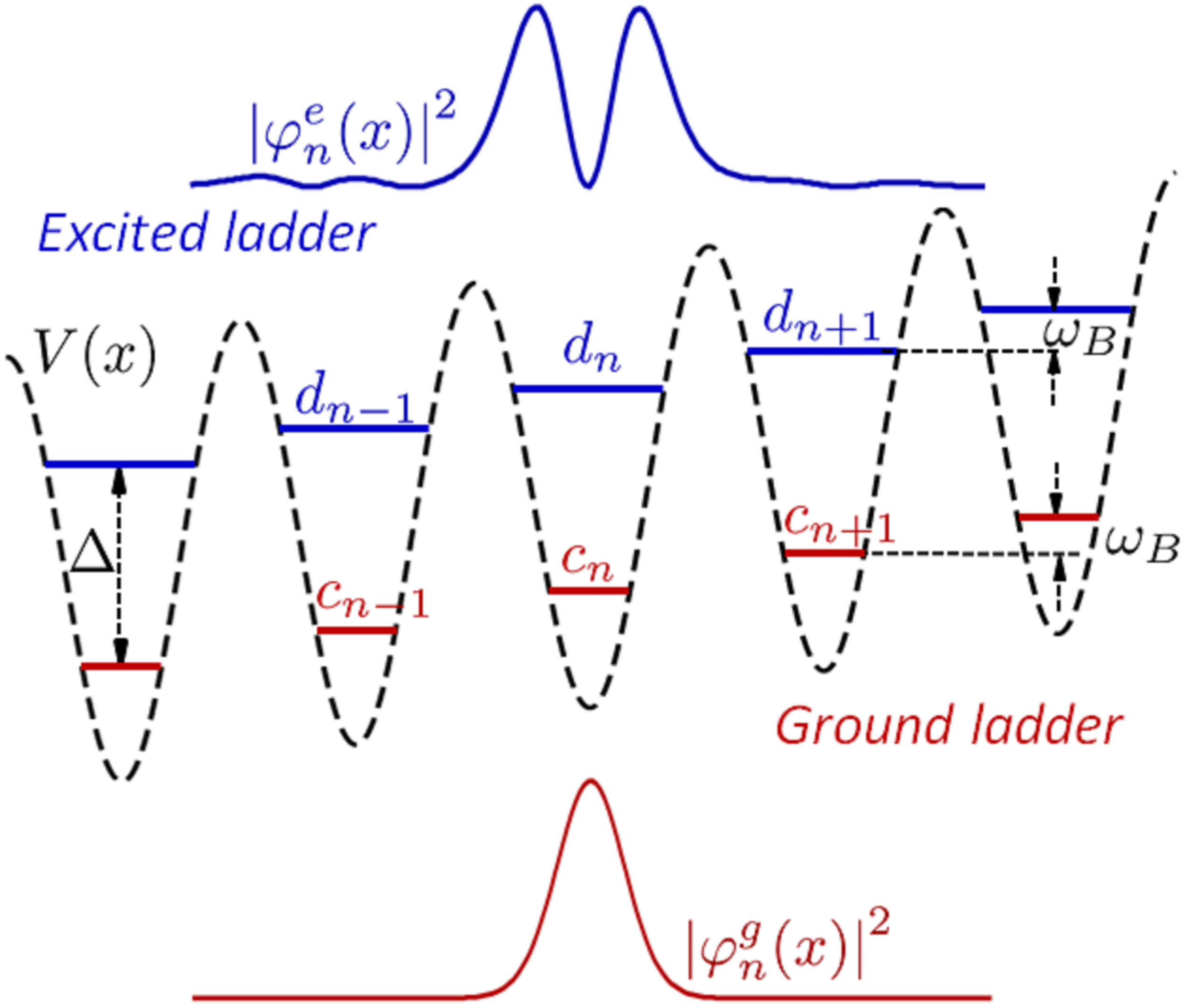} 
\par\end{centering}
\caption{\label{fig:WSstates}The Wannier-Stark system. Red energy levels $E_{n}^{g}$
and amplitude $c_{n}$ form the ``ground'' Wannier-Stark ladder,
the corresponding spatial probability distribution $\left|\varphi_{n}^{g}(x)\right|^{2}$
(for site $n$) is shown as the bottom red curve. Blue levels of energy
$E_{n}^{e}$ and amplitude $d_{n}$ form the ``excited'' WS ladder
and the corresponding eigenstate $\left|\varphi_{n}^{e}(x)\right|^{2}$
is shown as the top blue curve. Levels in the same well are separated
by an energy $\Delta$ and levels in the same ladder are separated
by $\omega_{B}$, the Bloch frequency. The parameters used in this
work are $V_{1}=6$, $F=1$, for which one finds numerically $\Delta=5.66$.}
\end{figure}

The present work combines these two driving forces in the ultracold-atom
field. We propose a general method for simulating Dirac physics in
a ``tilted'' one-dimensional optical lattice, a system that has
been very useful since the early days of the quantum simulation (even
before the term \emph{quantum simulation} was introduced), for example
for the observation of Bloch oscillations or the (equivalent) Wannier-Stark
ladder~\cite{BenDahan:BlochOsc:PRL96,Niu:LandauZennerWS:PRL96,Kolovsky:BECsOnTiltedLattices:PRA10,Kolovsky:BlochOscillationsBECDynamicalInst:Quantum:PRA09,Korsch:FractStab:EL00,Korsch:LifetimeWS:PRL99,Korsh:WSREV:IDiew:PREP02}.
The realization of such a system can be obtained by applying a far-detuned
laser standing wave that ultracold atoms see as a sinusoidal potential
acting on their center of mass variables~\citep{Cohen-TannoudjiDGO:AdvancesInAtomicPhysics::11}.
If the atom's de Broglie wavelength is comparable to the lattice constant
$\mathsf{a}=\lambda_{L}/2$, where $\lambda_{L}=2\pi/k_{L}$ is the
radiation wavelength (we use sans serif symbols for dimensioned quantities),
the system is in the quantum regime, a condition easily realized for
temperatures of the order of a few $\mu$K. In order to obtain a tilted
potential, one can simply chirp one of the beams forming the standing
wave: A linear shift of the frequency produces a quadratic displacement
of the nodes of the standing wave; in the rest frame with respect
to the nodes, an inertial constant force creates a tilt, that is,
a potential of the form $\mathsf{V}_{ws}(\mathsf{x})=-\mathsf{V}_{1}\cos(2k_{L}\mathit{\mathsf{x}})+\mathsf{Fx}$,
with $\mathsf{V}_{1}$ proportional to the radiation intensity and
$\mathsf{F}$ (constant) proportional to the frequency chirp. This
kind of setup is by now quite common in cold atom physics. In what
follows, we shall use dimensionless units such that spatial coordinate
$x=\mathsf{x}/\mathsf{a}$ is measured in units of the lattice potential
step $\mathsf{a}$, energy in units of the so-called ``recoil energy''
$\mathsf{E}_{R}=\hbar^{2}k_{L}^{2}/2M$ ($M$ is the mass of the atom),
time in units of $\hbar/\mathsf{E}_{R}$; $m^{*}=\pi^{2}/2$ is a
reduced mass, and $\hbar=1$ is the reduced Planck constant ~\citep{Thommen:WannierStark:PRA02}.
This defines the (dimensionless) \emph{Wannier-Stark} Hamiltonian
\begin{eqnarray}
H_{0} & = & \frac{p_{x}^{2}}{2m^{*}}-V_{1}\cos(2\pi x)+Fx,\label{eq:H0}
\end{eqnarray}
with $F\equiv\mathsf{Fa}/\mathsf{E}_{R}$ and $V_{1}=\mathsf{V_{1}}/\mathsf{E}_{R}$
. A given well (labeled by its position $x=n$) may, depending on
$V_{1}$ and $F$, host a number of bound eigenstates, called Wannier-Stark
(WS) states~\cite{Nenciu:WS:RMP91}. We note $\varphi_{n}^{\ell}(x)$
the $\ell^{\mathrm{th}}$ bounded state of well $n$~\footnote{Technically speaking, in infinite space, WS states are ``resonances''
\textendash{} metastable states~\cite{Nenciu:WS:RMP91}, but for
our present purposes they can be considered as stationary states as
long as the duration of the experiment is much shorter than their
lifetime. We checked numerically the validity of this hypothesis throughout
this work.} (see Fig.~\ref{fig:WSstates}), with the corresponding eigenenergy
$E_{n}^{\ell}$. The WS potential $V_{ws}$$=-V_{1}\cos(2\pi x)+Fx$
of Eq.~(\ref{eq:H0}) is invariant under a\emph{ simultaneous} spatial
translation by an integer multiple $m$ of the lattice constant $a=1$
\emph{and }an energy shift of $mF$, implying that $\varphi_{n+m}^{\ell}(x)=\varphi_{n}^{\ell}(x-m)$
and $E_{n+m}^{\ell}=E_{n}^{\ell}+m\omega_{B}$. These eigenenergies
form the so-called \emph{Wannier-Stark ladder} of step $\omega_{B}=F$,
called \emph{Bloch frequency} ($=|\mathsf{F}|\mathsf{a}/\hbar$ in
dimensioned units). In the present work we shall consider at most
two such ladders: The ground ladder $\ell=g$ of lowest energy and
the first excited ladder $\ell=e$.

A perturbation (for example a temporal or spatial modulation of $V_{1}$
or $\mathit{F}$), creates couplings between WS states and may generate
interesting dynamics~\citep{Thommen:WannierStark:PRA02,Thommen:DirectedWavePacket2D:PRA11,Zenesini:LandauZener:PRL09,Kolovsky:BECsOnTiltedLattices:PRA10,Goldman:PeriodicallyDrivenQuantumMatter:PRA15}.
The aim of the present work is to take advantage of these possibilities
to quantum-simulate Dirac dynamics. By an adequate choice of these
temporal modulations one can obtain either a spinor-2 model or a spinor-4
Dirac equation.

After a brief summary of the Dirac equation in sec.~\ref{sec:The-Dirac-equation},
sec.~\ref{sec:GeneralModel} introduces the general frame of our
study; the spinor-2 model and spinor-4 models are described in sec.~\ref{sec:Spinor-2}
and in sec.~\ref{sec:Spinor-4} respectively. Section~\ref{sec:ExperimentalRelization}
discusses the experimental feasibility of our theoretical proposals
and Sec.~\ref{sec:Conclusion} draws general conclusions of this
work.

Compared to other works demonstrating ways to simulate Dirac physics,
an advantage of our method is its simplicity both from the experimental
and the theoretical point of view. We use simple 1D optical lattices
modulated in time, for which analytic calculations can be pushed quite
far. The system is realizable experimentally with state-of-the-art
techniques (see Sec.~\ref{sec:ExperimentalRelization}). In particular,
no Raman or Zeeman transitions are necessary. Moreover, the approach
developed here is general and can be easily adapted to different situations,
as it will be seen below (and in future works).

\section{\label{sec:The-Dirac-equation}The Dirac equation in a nutshell}

The Dirac equation governs massive spin-1/2 particles~\citep{Dirac:QuantumTheoryElectron:PRSLA28,Pal:DiracMajoranaWeylFermions:AJP11}.
As shown by Dirac, the requirement for relativistic invariance leads
to the existence of spin and antiparticles; the theory deals with
a \emph{spinor-4}, that is, a 4-component state vector whose components
are themselves wave functions: 
\[
\boldsymbol{\text{\ensuremath{\psi}}}=\left(\begin{array}{c}
\psi_{1}(x,t)\\
\psi_{2}(x,t)\\
\psi_{3}(x,t)\\
\psi_{4}(x,t)
\end{array}\right).
\]
A possible representation for the Dirac equation for free particles
of mass $m$ is $H\boldsymbol{\text{\ensuremath{\psi}}}=i\partial_{t}\boldsymbol{\text{\ensuremath{\psi}}}$,
with the Dirac Hamiltonian 
\begin{equation}
H=\left(\boldsymbol{\alpha}\cdot\boldsymbol{p}c+\beta mc^{2}\right)\label{eq:H_Dirac}
\end{equation}
where $\alpha_{j}$ ($j=x,y,z$) and $\beta$ are Dirac matrices 
\[
\alpha_{j}=\left(\begin{array}{cc}
0 & \sigma_{j}\\
\sigma_{j} & 0
\end{array}\right),\qquad\beta=\left(\begin{array}{cc}
\mathbf{1} & 0\\
0 & -\mathbf{1}
\end{array}\right)
\]
with $\sigma_{j}$ the Pauli matrices, $\mathbf{1}$ the $2\times2$
identity matrix, $p_{j}=-i\partial/\partial x_{j}$ ($x_{j}=x,y,z$)
the momentum operator, $c$ the velocity of light, and $\hbar=1$.
For massive particles, in the rest frame of reference, the two upper
components of the spinor-4 can be identified with the spin components
of the (positive rest energy state) ``particle'' and the two bottom
components with the spin of the ``antiparticle'' (negative rest
energy state), but in a frame in which the particle is in motion,
the components are mixed and no such distinction is possible; a spinor-4
description is necessary. However this ``contamination'' is small
if $p\ll mc$. The general eigenvalues of the Dirac Hamiltonian are
$\pm\left(p^{2}c^{2}+m^{2}c^{4}\right)^{1/2}$, the distinction between
positive and negative eigenstates thus subsists (for a free particle)
in all cases.

For a massive free particle, if the momentum is parallel to the spin,
that is in the $z$ direction (the arbitrary quantization axis for
the spin), then the Dirac equation couples $\psi_{1}$ to $\psi_{3}$
and $\psi_{2}$ to $\psi_{4}$. If the momentum is orthogonal to the
spin (i.e. along the $x$- or the $y$-axis), it couples $\psi_{1}$
to $\psi_{4}$ and $\psi_{2}$ to $\psi_{3}$. Therefore, in both
cases the quantum dynamics can be described by two spinor-2, obeying
decoupled, equivalent equations. We can thus, for instance in the
latter case, form the \emph{spinor-2} 
\[
\bar{\boldsymbol{\psi}}=\left(\begin{array}{c}
\psi_{2}\\
\psi_{3}
\end{array}\right)
\]
which, from Eq.~(\ref{eq:H_Dirac}), obeys the spinor-2 Dirac equation

\begin{equation}
i\partial_{t}\bar{\boldsymbol{\psi}}=c\sigma_{j}p_{j}\bar{\boldsymbol{\psi}}+mc^{2}\sigma_{z}\bar{\boldsymbol{\psi}}\label{eq:DiracSpinor2}
\end{equation}
where $j=x$ or $j=y$. A similar equation holds for $(\psi_{1},\psi_{4})$.
In presence of a magnetic field, however, the quantization axis is
imposed by the field and for an arbitrary direction of the momentum
$\boldsymbol{p}$, the four components are coupled and the particle
is described by a true spinor-4.

Equation~(\ref{eq:H_Dirac}) is the original Hamiltonian written
by Dirac. This representation is well adapted to the case $p\ll mc$,
where the first term is small compared to the second; if the first
term is neglected, the Hamiltonian is diagonal. Other representations
exist, e.g., the so-called \emph{Weyl representation} corresponds
to the Hamiltonian 
\begin{equation}
H_{W}=c\left(\begin{array}{cc}
\boldsymbol{\sigma}\cdot\boldsymbol{p} & 0\\
0 & -\boldsymbol{\sigma}\cdot\boldsymbol{p}
\end{array}\right)+\gamma_{0}mc^{2}.\label{eq:H_Weyl}
\end{equation}
with

\[
\gamma_{0}=\left(\begin{array}{cc}
0 & \mathbf{1}\\
\mathbf{1} & 0
\end{array}\right).
\]
This representation is well suited for the ultra-relativistic limit
$p\gg mc$, where the mass term $\gamma_{0}mc^{2}$ in Eq.~(\ref{eq:H_Weyl})
becomes much smaller than the first one; neglecting the mass term
leaves a diagonal form. For massless particles, the system separates
into two subsets of equivalent equations, and can be described by
a spinor-2, the so-called \emph{Weyl fermion}. The above form implies
that these particles are characterized by a well-defined projection
of the spin along the particle's momentum $\boldsymbol{\sigma}\cdot\boldsymbol{p}/|\boldsymbol{p}|$,
a quantity called, as for photons, \emph{helicity}.

\section{\label{sec:GeneralModel}General model}

In this section we introduce the general model leading from Wannier-Stark
Hamiltonians of the form Eq.~(\ref{eq:H0}) to Dirac-like Hamiltonians.
We shall consider a restricted state space of one or two ladders,
i.e one or two WS states per potential well; the ground WS state (indexed
by $\ell=g$) $\varphi_{n}^{g}(x)=\left\langle x\right.\left|\varphi_{n}^{g}\right\rangle $
in the well $n$, of energy $E_{n}^{g}=n\omega_{B}$, and the first
excited WS state ($\ell=e$) $\varphi_{n}^{e}(x)$ of energy $E_{n}^{e}=E_{n}^{g}+\Delta$$=n\omega_{B}+\Delta$
of same well $n$ where $\varDelta$ is the energy offset between
$g$ and $e$ levels in the same well (cf.~Fig.~\ref{fig:WSstates}).
We assume in the following that none of these eigenenergies are degenerate.

The general evolution of an arbitrary wave function can then be written
in the form 
\begin{align}
\Psi(x,t)= & \sum_{n}\left[c_{n}(t)\exp\left(-iE_{n}^{g}t\right)\varphi_{n}^{g}(x)\right.\nonumber \\
 & \left.+d_{n}(t)\exp\left(-iE_{n}^{e}t\right)\varphi_{n}^{e}(x)\right]\label{eq:Psi-general}
\end{align}
with $c_{n}(0)=\left\langle \varphi_{n}^{g}\right.\left|\Psi(0)\right\rangle $
and $d_{n}(0)=\left\langle \varphi_{n}^{e}\right.\left|\Psi(0)\right\rangle $.

\begin{figure}
\begin{centering}
\includegraphics[width=0.9\columnwidth]{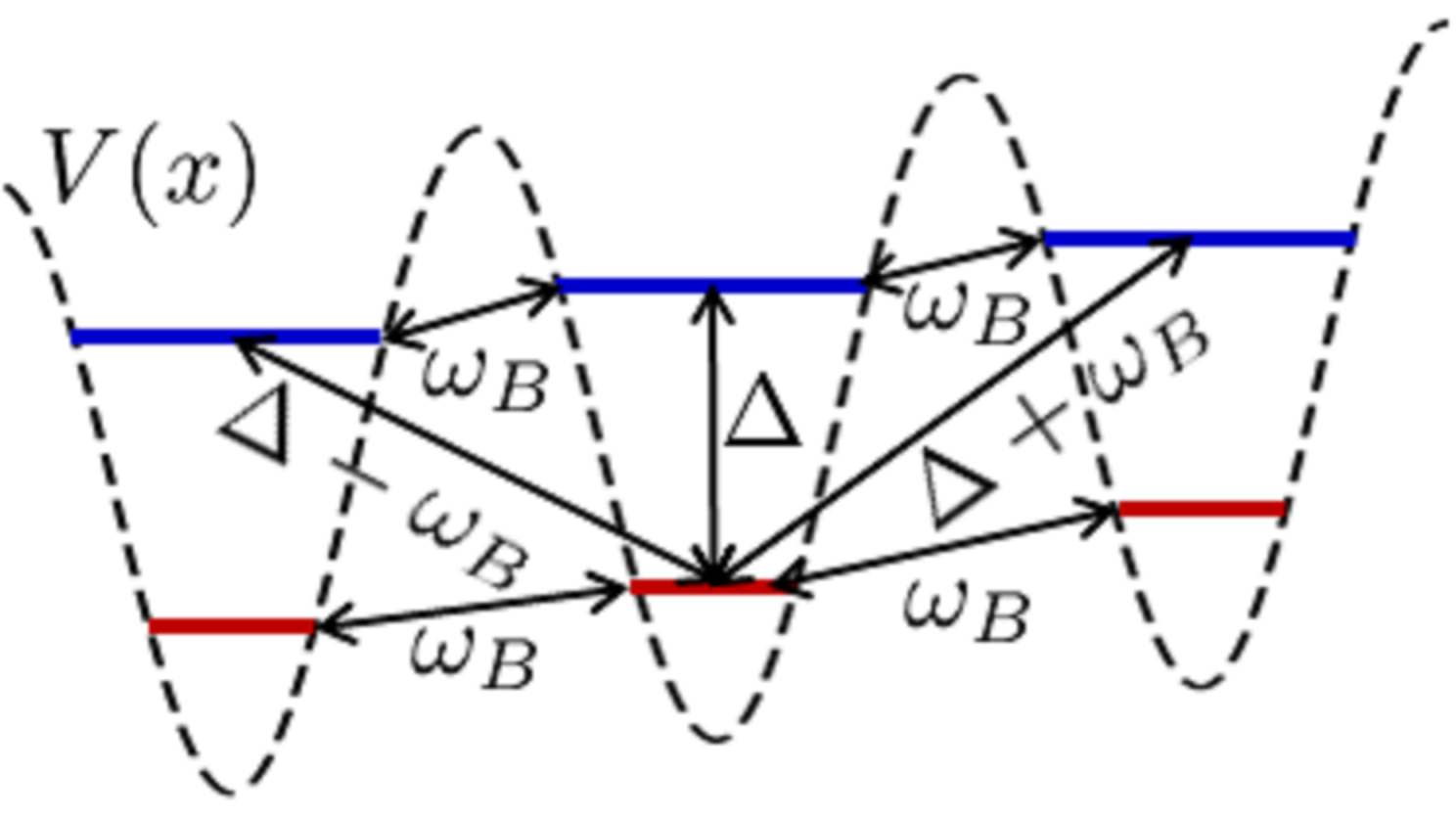} 
\par\end{centering}
\caption{\label{fig:WScouplings}Energy levels and couplings in the Wannier-Stark
system. A modulation of frequency $\omega_{B}$ induces an\emph{ intra-ladder}
coupling between adjacent wells. \emph{Inter-ladder} couplings are
induced by perturbation frequencies $\Delta-\omega_{B}$ ($n\rightarrow n-1$),
$\Delta$ ($n\rightarrow n$), and $\Delta+\omega_{B}$ ($n\rightarrow n+1$).}
\end{figure}

We introduce a perturbation $\bar{H}(t)$ so that our complete Hamiltonian
becomes $H=H_{0}+\bar{H}(t)$, with 
\begin{equation}
\bar{H}(t)=-V_{1}\cos(2\pi x)f_{1}(t)+V_{2}\cos(\pi x)f_{2}(t)+V_{S}(x).\label{eq:Hoverbar}
\end{equation}
A suitable choice of the frequencies present in $f_{1}(t)$ and $f_{2}(t)$
induces interactions between\textit{\emph{ states that are resonantly
coupled,}} as shown in Fig.~\ref{fig:WScouplings}. For example,
the ground-ladder level $\left|\varphi_{n}^{g}\right\rangle $ is
resonantly coupled to excited-ladder level $\left|\varphi_{n+1}^{e}\right\rangle $
by a modulation of frequency $\Delta+\omega_{B}$, and to $\left|\varphi_{n-1}^{e}\right\rangle $
by a modulation of frequency $\Delta-\omega_{B}$, and so on. The
perturbation term $V_{2}(x,t)=V_{2}\cos(\pi x)f_{2}(t)$ has double
spatial period, and $V_{S}(x)$ is a static contribution whose utility
will appear below.

Under the action of $\bar{H}$ the coupled equations of motion for
the amplitudes $c_{n}$ and $d_{n}$ of Eq.~(\ref{eq:Psi-general})
are developed in App.~\ref{sec:Derivation} and have the form: 
\begin{align}
i\frac{d}{dt}c_{n}= & \sum_{r\in\mathbb{Z}}\left\{ \left\langle \varphi_{n}^{g}\right|\bar{H}\left|\varphi_{n+r}^{g}\right\rangle e^{-ir\omega_{B}t}c_{n+r}\right.\nonumber \\
 & \left.+\left\langle \varphi_{n}^{g}\right|\bar{H}\left|\varphi_{n+r}^{e}\right\rangle e^{-ir\omega_{B}t}e^{-i\Delta t}d_{n+r}\right\} \nonumber \\
i\frac{d}{dt}d_{n}= & \sum_{r\in\mathbb{Z}}\left\{ \left\langle \varphi_{n}^{e}\right|\bar{H}\left|\varphi_{n+r}^{g}\right\rangle e^{-ir\omega_{B}t}e^{i\Delta t}c_{n+r}\right.\nonumber \\
 & \left.+\left\langle \varphi_{n}^{e}\right|\bar{H}\left|\varphi_{n+r}^{e}\right\rangle e^{-ir\omega_{B}t}d_{n+r}\right\} .\label{eq:GeneralEvolutionEqs}
\end{align}
The functions $f_{\alpha}(t)$ ($\alpha=1,2$) appearing in $\bar{H}$,
contain modulation frequencies of the form $\omega_{j,q}=j\omega_{B}+q\triangle$
with $j\in\mathbb{Z}$ and $q=0,\pm1$ 
\begin{equation}
f_{\alpha}(t)=\sum_{j,q}\left(A_{j,q}^{(\alpha)}e^{ij\omega_{B}t}e^{iq\Delta t}\right)\label{eq:f(t)}
\end{equation}
where the reality condition implies $A_{j,q}^{(\alpha)}=A_{-j,-q}^{(\alpha)*}$.
A great advantage of the Wannier-Stark model, within the assumption
that parameters are such that there are no intrinsically degenerated
states, is that tuning the amplitudes $A_{j,q}^{(\alpha)}$ allows
us to choose which pairs of states are coupled, providing a very flexible
control of the dynamics. For instance, one sees that modulations with
$q=0$ induce intra-ladder couplings ($g-g$ and $e-e$) and modulations
with $q=\pm1$ induce inter-ladder couplings $e-g$; taking\textcolor{blue}{{}
}$j=0$ creates a coupling $g-e$ in the \emph{same} well, whereas
$j=1$ couples wells $n\rightarrow n+1$ and $j=-1$ couples $n\rightarrow n-1$.

In the resonant case, Eqs.~(\ref{eq:GeneralEvolutionEqs}) can be
formally written as 
\begin{align}
i\frac{d}{dt}c_{n} & =\sum_{r}\left(T_{n,r}^{gg}c_{n+r}+T_{n,r}^{ge}d_{n+r}\right)\nonumber \\
i\frac{d}{dt}d_{n} & =\sum_{r}\left(T_{n,r}^{ee}d_{n+r}+T_{n,r}^{eg}c_{n+r}\right)\label{eq:general_discrete_model}
\end{align}
(see App.~\ref{sec:Derivation}). The explicit form of coupling coefficients
$T_{n,r}^{ab}$ ($a,b\in\left\{ e,g\right\} $) between the sites
$n$ and $n+r$ depend on the overlap integrals, which, thanks to
the properties of the WS states, are
\[
\left\langle \varphi_{n}^{g,e}\right|\cos(2\pi x)\left|\varphi_{n+r}^{g,e}\right\rangle =\left\langle \varphi_{0}^{g,e}\right|\cos(2\pi x)\left|\varphi_{r}^{g,e}\right\rangle ,
\]
\[
\left\langle \varphi_{n}^{g,e}\right|\cos(\pi x)\left|\varphi_{n+r}^{g,e}\right\rangle =(-1)^{n}\left\langle \varphi_{0}^{g,e}\right|\cos(\pi x)\left|\varphi_{r}^{g,e}\right\rangle .
\]
One then obtains intra-ladder coupling as 
\begin{align}
T_{n,r}^{gg}=\left\langle \varphi_{n}^{g}\right|V_{S}\left|\varphi_{n}^{g}\right\rangle \delta_{r,0}- & V_{1}A_{r,0}^{(1)}\left\langle \varphi_{0}^{g}\right|\cos(2\pi x)\left|\varphi_{r}^{g}\right\rangle \nonumber \\
 & +(-1)^{n}V_{2}A_{r,0}^{(2)}\left\langle \varphi_{0}^{g}\right|\cos(\pi x)\left|\varphi_{r}^{g}\right\rangle \nonumber \\
T_{n,r}^{ee}=\left\langle \varphi_{n}^{e}\right|V_{S}\left|\varphi_{n}^{e}\right\rangle \delta_{r,0}- & V_{1}A_{r,0}^{(1)}\left\langle \varphi_{0}^{e}\right|\cos(2\pi x)\left|\varphi_{r}^{e}\right\rangle \nonumber \\
 & +(-1)^{n}V_{2}A_{r,0}^{(2)}\left\langle \varphi_{0}^{e}\right|\cos(\pi x)\left|\varphi_{r}^{e}\right\rangle .\label{eq:interLadderCoupls}
\end{align}
and inter-ladder couplings 
\begin{align}
T_{n,r}^{ge}= & -V_{1}A_{r,1}^{(1)}\left\langle \varphi_{0}^{g}\right|\cos(2\pi x)\left|\varphi_{r}^{e}\right\rangle \nonumber \\
 & +(-1)^{n}V_{2}A_{r,1}^{(2)}\left\langle \varphi_{0}^{g}\right|\cos(\pi x)\left|\varphi_{r}^{e}\right\rangle \nonumber \\
T_{n,r}^{eg}= & -V_{1}A_{r,-1}^{(1)}\left\langle \varphi_{0}^{e}\right|\cos(2\pi x)\left|\varphi_{r}^{g}\right\rangle \nonumber \\
 & +(-1)^{n}V_{2}A_{r,-1}^{(2)}\left\langle \varphi_{0}^{e}\right|\cos(\pi x)\left|\varphi_{r}^{g}\right\rangle .\label{eq:intraLadderCoupls}
\end{align}

This general model spans all cases we will consider in the present
work. In Sec.~\ref{sec:Spinor-2} we show how to construct a quantum
simulator for a spinor-2 Dirac equation, and in Sec.~\ref{sec:Spinor-4}
we show how the full spinor-4 Dirac or Weyl equations can be synthesized.

\section{\label{sec:Spinor-2}Spinor-2 model}

Many interesting phenomena related to the Dirac equation can be illustrated
with a simpler spinor-2. In order to construct a spinor-2 quantum
simulator we restrict our system to the ground state ladder with ``self''
($c_{n}$$\leftrightarrows c_{n}$) and nearest neighbors ($c_{n}\leftrightarrows c_{n\pm1}$)
couplings. Inter-ladder transitions are set off by keeping only the
$q=0$ term in Eq.~(\ref{eq:f(t)}), and we start with an initial
condition $d_{n}(0)=0$ for all sites~\footnote{Experimentally this can be done by trapping the atoms on a shallow
optical lattice and increasing adiabatically the lattice amplitude
to the desired level.}, so that the excited ladder is never populated. We also set $V_{1}=V_{S}=0$
in Eq.~(\ref{eq:Hoverbar}). The perturbation thus contains only
contributions of double spatial period 
\begin{equation}
\bar{H}=V_{2}f_{2}(t)\cos(\pi x)\label{eq:Spinor2H1}
\end{equation}
with, in Eq.~(\ref{eq:f(t)}), $j=0,\pm1$, $q=0$, that is 
\begin{align}
f_{2} & (t)=A_{0,0}^{(2)}+A_{1,0}^{(2)}e^{i\omega_{B}t}+A_{-1,0}^{(2)}e^{-i\omega_{B}t}\nonumber \\
 & =A_{0}+A_{1}e^{i\omega_{B}t}+A_{1}^{*}e^{-i\omega_{B}t},\label{eq:Dirac2_modulation}
\end{align}
where, in the second line, we suppressed for simplicity the fixed
indexes $q=0$ and $\alpha=2$. The remaining coupling parameters
are then {[}Eq.~(\ref{eq:interLadderCoupls}){]} 
\begin{align*}
T_{n,1}^{gg} & =(-1)^{n}V_{2}A_{1}\left\langle \varphi_{0}^{g}\right|\cos(\pi x)\left|\varphi_{1}^{g}\right\rangle \\
T_{n,0}^{gg} & =(-1)^{n}V_{2}A_{0}\left\langle \varphi_{0}^{g}\right|\cos(\pi x)\left|\varphi_{0}^{g}\right\rangle \\
T_{n,-1}^{gg} & =(-1)^{n}V_{2}A_{-1}\left\langle \varphi_{0}^{g}\right|\cos(\pi x)\left|\varphi_{-1}^{g}\right\rangle \\
 & =-(-1)^{n}V_{2}A_{1}^{*}\left\langle \varphi_{0}^{g}\right|\cos(\pi x)\left|\varphi_{1}^{g}\right\rangle .
\end{align*}
Eqs.~(\ref{eq:general_discrete_model}) then imply

\begin{align}
i\frac{d}{dt}c_{n}= & (-1)^{n}V_{2}A_{0}\left\langle \varphi_{0}^{g}\right|\cos(\pi x)\left|\varphi_{0}^{g}\right\rangle c_{n}\nonumber \\
 & (-1)^{n}V_{2}\left\langle \varphi_{0}^{g}\right|\cos(\pi x)\left|\varphi_{1}^{g}\right\rangle \left[A_{1}c_{n+1}-A_{1}^{*}c_{n-1}\right].\label{eq:Dirac2_discret}
\end{align}
A key point for realizing a spinor-2 system is that the perturbation
of double spatial period creates alternate sign couplings from site
to site (see App.~\ref{sec:Derivation}). This has a dynamical effect
that is clearly visible in the reciprocal space, where we define ``spin''
states as ``odd site'' and ``even site'' amplitudes 
\begin{eqnarray}
\bar{c}_{+}(k,t) & = & \sum_{n}e^{2ink}c_{2n}(t)\nonumber \\
\bar{c}_{-}(k,t) & = & \sum_{n}e^{i(2n+1)k}c_{2n+1}(t).\label{eq:creciproc}
\end{eqnarray}
Taking, for simplicity, $A_{1}$ real in Eq.~(\ref{eq:Dirac2_discret}),
one obtains the following coupled set of equations 
\begin{eqnarray}
i\frac{d}{dt}\bar{c}_{+}(k,t) & = & E_{0}\bar{c}_{+}(k,t)-2i\Omega_{2}\sin k\:\bar{c}_{-}(k,t)\nonumber \\
i\frac{d}{dt}\bar{c}_{-}(k,t) & = & -E_{0}\bar{c}_{-}(k,t)+2i\Omega_{2}\sin k\:\bar{c}_{+}(k,t),\label{eq:discretModel}
\end{eqnarray}
where we defined the frequency $\Omega_{2}=$$V_{2}A_{1}\left\langle \varphi_{0}^{g}\right|\cos(\pi x)\left|\varphi_{1}^{g}\right\rangle $
and the ``self-energy'' $E_{0}=$$V_{2}A_{0}\left\langle \varphi_{0}^{g}\right|\cos(\pi x)\left|\varphi_{0}^{g}\right\rangle $.
These two equations show the emergence of an effective pseudo spinor-2
which in $k$-space is
\[
\bar{\boldsymbol{\psi}}=\left(\begin{array}{c}
c_{+}(k,t)\\
c_{-}(k,t)
\end{array}\right).
\]
 Looking for solutions in $\exp\left(-i\omega(k)t\right)$, the corresponding
eigenenergies $\omega(k)$ are

\begin{equation}
\omega_{\pm}(k)=\pm\sqrt{E_{0}^{2}+4\Omega_{2}^{2}\sin^{2}k}.\label{eq:2bands}
\end{equation}

For $E_{0}=0$, the positive and negative eigenenergies $\pm2\left|\Omega_{2}\sin k\right|$
are associated to the eigenspinor 
\begin{equation}
\bar{\boldsymbol{\psi}}_{\pm}=\frac{1}{\sqrt{2}}\left(\begin{array}{c}
1\\
\pm i\mathrm{sgn}(\Omega_{2}k)
\end{array}\right),\label{eq:Spinor2Eigenstates}
\end{equation}
where $\mathrm{sgn}(x)$ is the sign function. The linear, phonon-like,
dispersion relation for $k\rightarrow0$ , $\omega_{\pm}(k)=\pm2\left|\Omega_{2}k\right|$,
reproduces the spectrum of the relativistic massless spin-1/2 fermion.
A ``1D-conical intersection'' occurs as the two branches coalesce
at $k=0$, creating a so-called \emph{Dirac point}.

In real space, if the even- $c_{2n}(x,t)$ and odd-site $c_{2n+1}(x,t)$
amplitudes vary slowly on the scale of the lattice step $a=1$, one
can take the continuous limit of Eqs.~(\ref{eq:Dirac2_discret}),
and define the functions $c_{\pm}(x,t)$ as the spatial envelopes
of the $c_{n}(x,t)$ (cf. App.~\ref{sec:Derivation}), leading to
the spinor-2

\[
\boldsymbol{\phi}=\left(\begin{array}{c}
c_{+}(x,t)\\
c_{-}(x,t)
\end{array}\right)
\]
which obeys an equation 
\begin{equation}
i\partial_{t}\boldsymbol{\phi}=-2\Omega_{2}(-i\partial_{x})\sigma_{y}\boldsymbol{\phi}+E_{0}\sigma_{z}\boldsymbol{\phi}\label{eq:SyntheticDirac2}
\end{equation}
of the same form as Eq.~(\ref{eq:DiracSpinor2}) if one sets $p_{y}=-i\partial_{x}$
(the labeling of the axes is obviously arbitrary). By comparing Eqs.~(\ref{eq:SyntheticDirac2})
and~(\ref{eq:H_Dirac}) we can make the following identifications:
$E_{0}=\overline{m}\overline{c}^{2}$ and $2\left|\Omega_{2}\right|=\overline{c}$,
where $\overline{m}$ and $\overline{c}$ are the effective mass and
speed of light which can be adjusted by changing the modulation amplitudes
$A_{0}$ and $A_{1}$ in Eq.~(\ref{eq:Dirac2_modulation}).

The validity of the model Eq.~(\ref{eq:discretModel}) can be numerically
tested by comparison with the simulation of the exact Schr\"odinger
equation corresponding to the Hamiltonian $H_{0}+\bar{H}$ with $\bar{H}$
given by Eq.~(\ref{eq:Spinor2H1}). We chose a broad initial wave
packet, with amplitudes: 
\begin{align}
c_{2n} & =a_{+}G^{(k_{0})}(2n),\;c_{2n+1}\label{eq:CI}\\
 & =a_{-}G^{(k_{0})}(2n+1),
\end{align}
with $G^{(k_{0})}(n)=\left(2\pi/\sigma\right)^{1/2}$$\exp\left(-ink_{0}\right)$$\exp\left(-n^{2}/\sigma^{2}\right)$,
$\sigma\gg1,$ with the normalization condition $\left|a_{+}\right|^{2}+\left|a_{-}\right|^{2}=1$.
The initial spinor is thus 
\begin{equation}
\boldsymbol{\phi}_{0}=\left(\begin{array}{c}
a_{\text{+}}\\
a_{-}
\end{array}\right)G^{(k_{0})}(x)\leftrightarrow\left(\begin{array}{c}
a_{\text{+}}\\
a_{-}
\end{array}\right)\bar{G}^{(k_{0})}(k)\label{eq:CI_x}
\end{equation}
where the first expression is in real and the second in momentum space,
and $\bar{G}^{(k_{0})}(k)$ is a narrow Gaussian function centered
at $k=k_{0}$.

\begin{figure}[th]
\begin{centering}
\includegraphics[width=7cm]{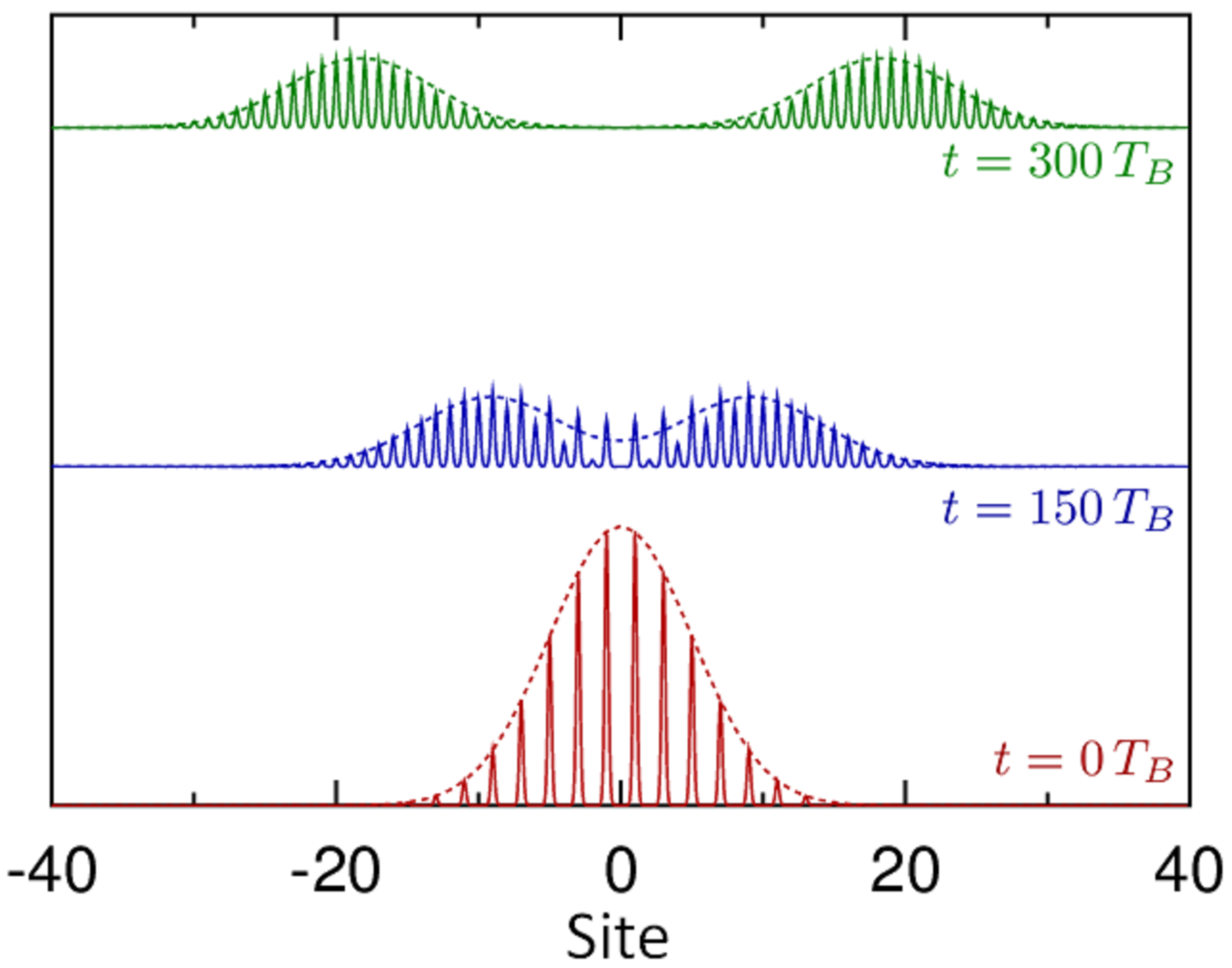} 
\par\end{centering}
\caption{Evolution of an initial wave packet with $k_{0}=0$, $\sigma=10$
and $(a_{+},a_{-})=(1,0)$ {[}Eq.~(\ref{eq:CI}){]} using the discrete
model given by Eq.~(\ref{eq:Dirac2_discret}) (dashed lines), at
times $t=0$ (red bottom line), $150T_{B}$ (blue middle line) and
$300T_{B}$ (top green line) and compared to the \textcolor{black}{exact
Schr}\"o\textcolor{black}{dinger equation simulation (full lines).}
Parameters are $V_{1}=6$, $F=1,$ $\bar{H}(x,t)=0.5\cos\left(\pi x\right)\cos\left(\omega_{B}t\right)$
which give $\left\langle \varphi_{0}^{g}\right|\cos(\pi x)\left|\varphi_{1}^{g}\right\rangle =-2.31\times10^{-2}$
(numerical value), and $\Omega_{2}=-5.4\times10^{-3}$. The numerical
value of the drift velocity $v_{D}$ agrees with the theoretical value
$v_{D}=2\left|\Omega_{2}\cos k_{0}\right|=1.1\times10^{-2}$\textcolor{black}{. }}

\label{fig:dyn_behavior} 
\end{figure}

The dashed lines in Fig.~\ref{fig:dyn_behavior} show the dynamical
behavior of a massless particle (setting $A_{0}=0$ in Eq.~(\ref{eq:Dirac2_modulation})
leads to $E_{0}=0$) obtained from the above model {[}cf.~Eq.~(\ref{eq:Dirac2_discret}){]}
at time $t=0,$ $150T_{B}$ and $300T_{B}$, where $T_{B}=2\pi/F$
is the Bloch-period. The initial spinor $(a_{+},a_{-})=(1,0)$ with
$k_{0}\rightarrow0^{+}$, corresponds to a superposition of the positive
energy eigenspinor $(1,i)/\sqrt{2}$ and the negative energy eigenspinor
$(1,-i)/\sqrt{2}$ {[}cf.~Eq.~(\ref{eq:Spinor2Eigenstates}){]}
having opposite drift velocities $\pm v_{D}$ which from Eq.~(\ref{eq:2bands})
read
\[
v_{D}=\left|\frac{d\omega_{\pm}}{dk}\right|_{k_{0}}=2\left|\Omega_{2}\cos k_{0}\right|\simeq2\left|\Omega_{2}\right|.
\]
The comparison with the solution of exact Schr\"odinger equation
(full lines) shows a very good agreement up to $t=300T_{B}$. One
can verify that splitting into two separate wave packets moving with
opposite group velocities $\pm v_{D}$ which matches the expected
theoretical value.

\begin{figure}[h]
\begin{centering}
\includegraphics[width=0.9\columnwidth]{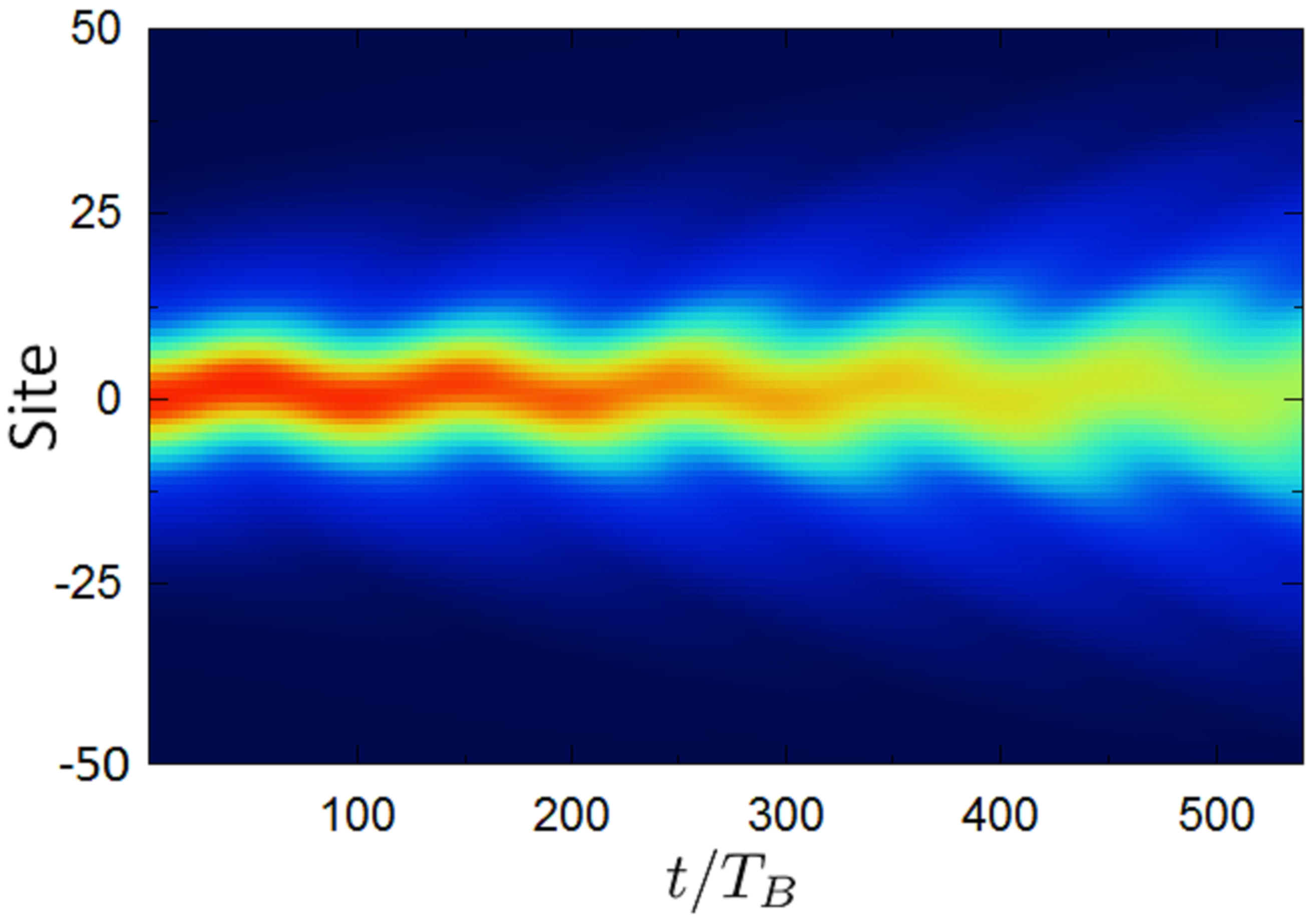} 
\par\end{centering}
\caption{\emph{Zitterbewegung} obtained from the discrete model Eq.~(\ref{eq:Dirac2_discret})
with initial spinor $(a_{+},a_{-})=2^{-1/2}(1,1)$, $\sigma=10$ and
$k_{0}=0$. The probability density is displayed in false colors.
Potential parameters are the same as in Fig.~\ref{fig:dyn_behavior}
except $\bar{H}(x,t)=\cos\pi x\left(0.5\cos\omega_{B}t+0.005\right)$.
The time-independent contribution $A_{0}=0.005$ in $\bar{H}$ leads
to a mass term $E_{0}=4.6\times10^{-3}$.}

\label{fig:ZBFalseColors} 
\end{figure}

One of the most characteristic effects associated to the Dirac equation
for \emph{massive} particles is the so-called \emph{Zitterbewegung}
(``trembling motion''), an interference effect between the positive
and negative energy parts of the spinor resulting in a spatial jitter
of the wave packet~\cite{Vaishnav:ObservingZitterbewegungUltracold:PRL08}.
Such an effect was recently observed in quantum simulators of the
Dirac equation with trapped ions~\citep{Gerritsma:QuantumSimulationDirac:N10},
with ultracold atoms~\cite{LeBlanc:ObservationZitterbewegungBEC:NJP13,Qu:ObservationZitterbewegungBEC:PRA13},
and in a photonic device~\cite{Dreisow:ClassicalSimulationZitterbewegungPhotLatt:PRL10,Longhi:PhotonicAnalogZitterbewegung:OL10}.
Figure~\ref{fig:ZBFalseColors} shows the spatio-temporal behavior
of a wave packet for a massive particle governed by Eqs.~(\ref{eq:Dirac2_discret}),
with a an initial spinor $(a_{+},a_{-})=2^{-1/2}(1,1)$ and $k_{0}=0$,
corresponding to superposition of positive and negative energy eigenstates
(as can be seen from Eq.~(\ref{eq:discretModel}) in the limit $k\rightarrow0$).
In order to give a mass to the particle, we set $A_{0}\neq0$ in Eq.~(\ref{eq:Dirac2_modulation}),
so that $E_{0}\neq0$). We verified that the same spatio-temporal
behavior is obtained from the exact Schr\"odinger equation.

\begin{figure}[h]
\begin{centering}
\includegraphics[width=0.9\columnwidth]{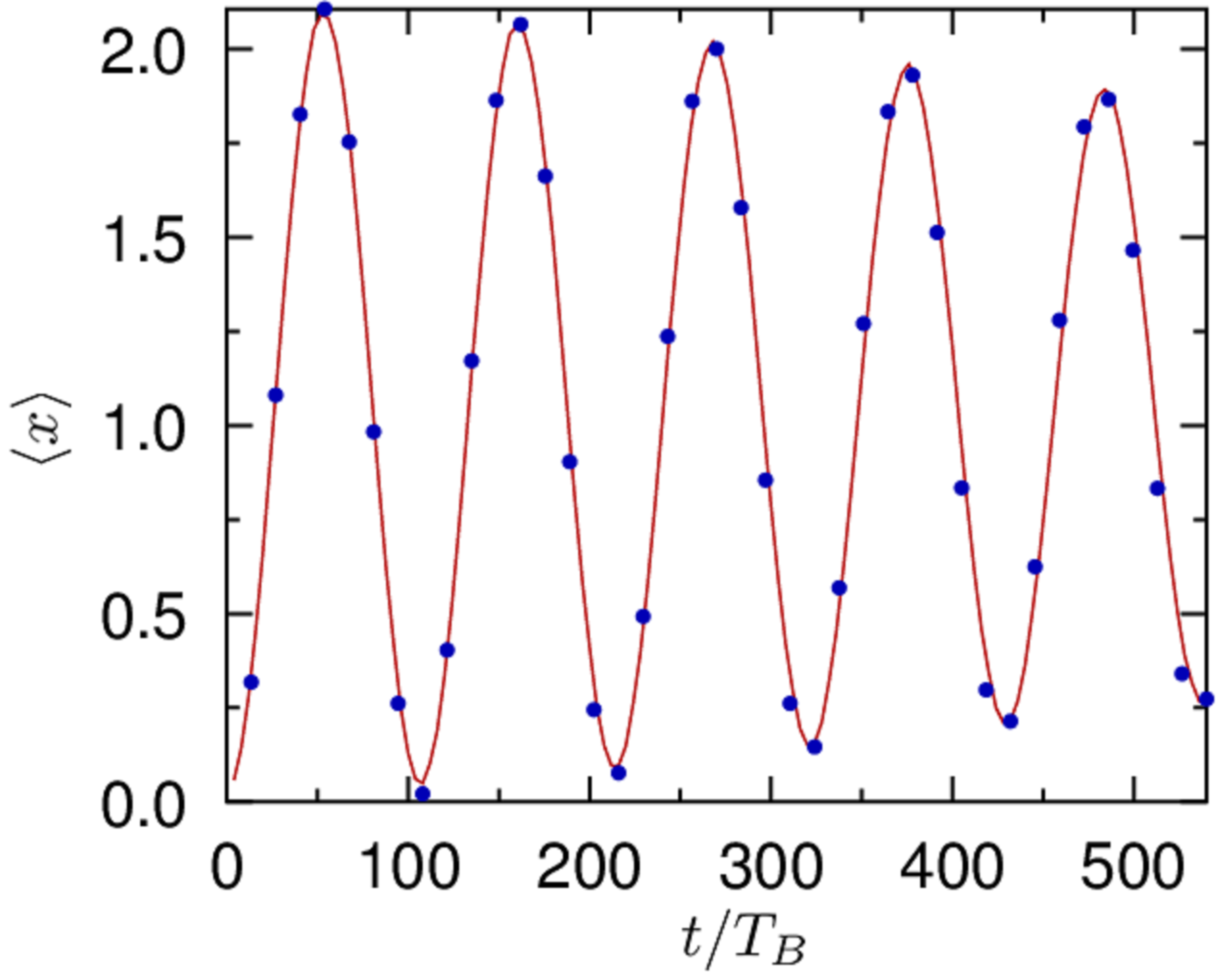} 
\par\end{centering}
\caption{Evolution of the average position $\left\langle x(t)\right\rangle $
for $0\leq t\leq5T_{ZB}$. Solid red line: Numerical result obtained
from the exact Schr\"odinger equation. Blue circles: calculation
from the discrete model Eq.~(\ref{eq:Dirac2_discret}). Same parameters
as in Fig.~\ref{fig:ZBFalseColors}. The Zitterbewegung period is
in excellent agreement with the theoretical value, $T_{ZB}=2\pi/(2E_{0})=109T_{B}$,
and its amplitude with the prediction $\left|\Omega_{2}\right|/E_{0}=1.17$.
Due to diffusion, the amplitude is attenuated by a factor $(1+D^{2}t^{2})^{-1/2}\sim0.75$
at $t=5T_{ZB}$ as compared to its initial value. }

\label{fig:ZB_discret} 
\end{figure}

From Eq.~(\ref{eq:discretModel}) one can obtain the evolution of
the wave packet's average position 
\begin{align*}
\frac{d\left\langle x\right\rangle }{dt} & =\frac{1}{i\hbar}\left\langle \left[x,H\right]\right\rangle =-2\Omega_{2}\left\langle \sigma_{y}\right\rangle \\
 & =2i\Omega_{2}\int dx\left(c_{+}^{*}(x,t)c_{-}(x,t)-\mathrm{c.c.}\right)\\
 & =2i\Omega_{2}\int dk\left(\bar{c}_{+}^{*}(k,t)\bar{c}_{-}(k,t)-\mathrm{c.c.}\right).
\end{align*}
The fact that the oscillation depends on $c_{+}^{*}c_{-}$ (in real
or momentum space) shows that the Zitterbewegung is due to the coherence
between positive- and negative-energy states, confirming its physical
interpretation as a quantum beat between odd- and even- site contributions
(or positive and negative energy states in Dirac's language). To the
leading order in $k\approx0$ we find 
\begin{equation}
\frac{d\left\langle x\right\rangle }{dt}=\frac{i2\Omega_{2}}{\sqrt{1-iDt}}a_{+}^{*}a_{-}e^{2iE_{0}t}+\mathrm{c.c}\label{eq:x_moyen}
\end{equation}
with $D=4\Omega_{2}^{2}/(E_{0}\sigma^{2})$. In this approximation,
the amplitude of the oscillation is seen to be directly proportional
to the initial coherence $a_{+}^{*}a_{-}$. The oscillation has frequency
$2E_{0}$, as it is the case for the electron's Zitterbewegung, and
is slowly damped by diffusion effects with an effective coefficient
$D$; note that the amplitude of the oscillation for $Dt\rightarrow0$,
is $\left|\Omega_{2}\right|/E_{0}=(2\bar{m}\overline{c})^{-1}$, that
is, half the dimensionless Compton wavelength, also in agreement with
the Zitterbewegung of an electron. As shown in Fig.~\ref{fig:ZB_discret},
the numerical calculations of $\left\langle x(t)\right\rangle $ from
the exact Schr\"odinger equation and from the discrete model are
in excellent agreement and match the theoretical amplitude and period
deduced from Eq.~(\ref{eq:x_moyen}).

The effective parameters $\overline{m}=E_{0}/4\Omega_{2}^{2}$ and
$\overline{c}=2\left|\Omega_{2}\right|$ can be calculated from the
parameters used in the above simulations. For atoms of mass $M$,
they read, in dimensioned units, $\overline{\mathsf{c}}=2\left|\Omega_{2}\right|E_{R}d/\hbar$$=\left|\Omega_{2}\right|\left(2\pi^{2}\hbar/M\lambda_{L}\right)$
and $\overline{\mathsf{m}}=E_{0}E_{R}/c^{2}$$=\left(E_{0}/2\pi^{2}\Omega_{2}^{2}\right)M$.
For cesium atoms and for potential parameters chosen in this section
(\emph{$E_{0}=4.6\times10^{-3}$, $\left|\Omega_{2}\right|=5.4\times10^{-3}$})
this leads to $\overline{\mathsf{m}}\sim9.3M$ and $\overline{\mathsf{c}}\sim1.33\times10^{-4}\left|\Omega_{2}\right|$$\sim7\times10^{-7}$m/s$\approx2\times10^{-3}v_{R}$,
where $v_{R}=\sqrt{2E_{R}/M}$ is the atom recoil velocity .

\section{\label{sec:Spinor-4}Spinor-4 model}

We can also construct a full Dirac equation with a spinor-4. Using
different coupling schemes we obtain either a Dirac-like equation
in the standard representation or its analog in the Weyl representation.
This beautifully illustrates the flexibility of the general model
presented in Sec.~\ref{sec:GeneralModel}.

\subsection{Spinor-4 Dirac representation }

In order to construct a spinor-4 in the Dirac representation, we consider
both ground and excited WS ladders, nearest-neighbors inter-ladder
couplings are set on and intra-ladder couplings are set off. The perturbation
is thus of the form {[}cf.~Eq.~(\ref{eq:Hoverbar}){]} 
\[
\bar{H}=-V_{1}f_{1}(t)\cos(2\pi x)+V_{S}(x)
\]
with the modulation function 
\[
f_{1}(t)=A_{1,1}^{(1)}e^{i\omega_{B}t}e^{i\Delta t}+A_{1,-1}^{(1)}e^{i\omega_{B}t}e^{-i\Delta t}+\mathrm{c.c}.
\]
From Eq.~(\ref{eq:general_discrete_model}) we obtain the equations
of motion 
\begin{eqnarray*}
i\frac{d}{dt}c_{n} & = & T_{n,0}^{gg}c_{n}+T_{n,1}^{ge}d_{n+1}+T_{n,-1}^{ge}d_{n-1}\\
i\frac{d}{dt}d_{n} & = & T_{n,0}^{ee}d_{n}+T_{n,1}^{eg}c_{n+1}+T_{n,-1}^{eg}c_{n-1}
\end{eqnarray*}
with 
\[
T_{n,-1}^{eg}=\left(T_{n,1}^{ge}\right)^{*},\;T_{n,1}^{eg}=\left(T_{n,-1}^{g,e}\right)^{*}.
\]
A Dirac-like equation is obtained if the coupling coefficients $T_{n,1}^{ge}=-T_{n,-1}^{ge}$
are imaginary and if $A_{1,1}^{(1)}\left\langle \varphi_{0}^{g}\right|\cos(2\pi x)\left|\varphi_{1}^{e}\right\rangle $$=-A_{1,-1}^{(1)}\left\langle \varphi_{1}^{g}\right|\cos(2\pi x)\left|\varphi_{0}^{e}\right\rangle $,
a condition that is realized by tuning the modulation amplitudes $A_{1,\pm1}^{(1)}$
so that they exactly compensate for the difference in the overlap
integrals. The static perturbation $V_{S}(x)$ is chosen to be translation-invariant
with respect to the reference lattice constant $a=1$, so that $T_{n,0}^{gg}=\left\langle \varphi_{0}^{g}\right|V_{S}\left|\varphi_{0}^{g}\right\rangle \equiv V_{S}^{g}$
and $T_{n,0}^{ee}=\left\langle \varphi_{0}^{e}\right|V_{S}\left|\varphi_{0}^{e}\right\rangle \equiv V_{S}^{e}$
do not depend on $n$; the simple form used here is $V_{S}(x)\propto\cos\left(4\pi x\right)$.
Thus 
\begin{eqnarray}
i\frac{d}{dt}c_{n} & = & E_{0}c_{n}+i\Omega_{1}\left(d_{n+1}-d_{n-1}\right)\nonumber \\
i\frac{d}{dt}d_{n} & = & -E_{0}d_{n}+i\Omega_{1}\left(c_{n+1}-c_{n-1}\right)\label{eq:Spinor4cndn}
\end{eqnarray}
where the coupling $\Omega_{1}$ is given by 
\[
i\Omega_{1}=T_{n,1}^{ge}=-V_{1}A_{1,1}^{(1)}\left\langle \varphi_{0}^{g}\right|\cos(2\pi x)\left|\varphi_{1}^{e}\right\rangle 
\]
(with $A_{1,1}^{(1)}$ imaginary) and the effective rest mass $E_{0}$,
controlled by the static potential $V_{S}(x),$ is given by~\footnote{We simply redefined $\left(c_{n},d_{n}\right)$ as $\left(c_{n},d_{n}\right)\exp\left[-i(V_{S}^{g}+V_{S}^{e})t/2\right]$.}
\begin{align}
E_{0} & =\frac{V_{S}^{g}-V_{S}^{e}}{2}.\label{eq:E_0_4spinor}
\end{align}

The coupled equations~(\ref{eq:Spinor4cndn}) can be split into two
independent sub-lattices corresponding to sites $c_{n}$ with $n$
even coupled to $d_{n}$ with $n$ odd and conversely. Hence, we can
build a 4-component Wannier-Stark spinor
\begin{equation}
\boldsymbol{\psi}=\left(\begin{array}{c}
c_{+}\\
c_{-}\\
d_{+}\\
d_{-}
\end{array}\right)\label{eq:spinor-4}
\end{equation}
where $c_{\pm}(x,t)$ and $d_{\pm}(x,t)$ are the slowly varying envelopes
of $c_{n}$ and $d_{n}$ for $n$ odd and $n$ even respectively (in
close analogy with what has been done in the spinor-2 case, Sec.~\ref{sec:Spinor-2}
and in App.~\ref{sec:Derivation}), giving 
\begin{equation}
i\partial_{t}\boldsymbol{\psi}=\left(E_{0}\beta-2\Omega_{1}\alpha_{x}p_{x}\right)\boldsymbol{\psi}\label{eq:Dirac-4}
\end{equation}
which corresponds to the Dirac equation described by Eq.~(\ref{eq:H_Dirac}).
As stated in Sec.~\ref{sec:The-Dirac-equation}, this equation can
be decoupled into two equivalent sets 
\[
i\frac{\partial}{\partial t}\left(\begin{array}{c}
c_{+}\\
d_{-}
\end{array}\right)=\left(E_{0}\sigma_{z}-2\Omega_{1}p_{x}\sigma_{x}\right)\left(\begin{array}{c}
c_{+}\\
d_{-}
\end{array}\right)
\]
the other components $(c_{-},d_{+})$ following exactly the same equation.
The corresponding dispersion relation is again $\omega_{\pm}(k)=\pm\left(E_{0}^{2}+4\Omega_{1}^{2}k^{2}\right)^{1/2}$,
but each eigenvalue has now a double degeneracy. Note that this degeneracy
can be lifted by adding other terms in $\bar{H}$ (for instance, terms
proportional to $\cos(\pi x)$ which break translation invariance
with respect to the lattice step $a=1$) and will be studied in a
forthcoming paper.

\begin{figure}[h]
\begin{centering}
\includegraphics[width=0.9\columnwidth]{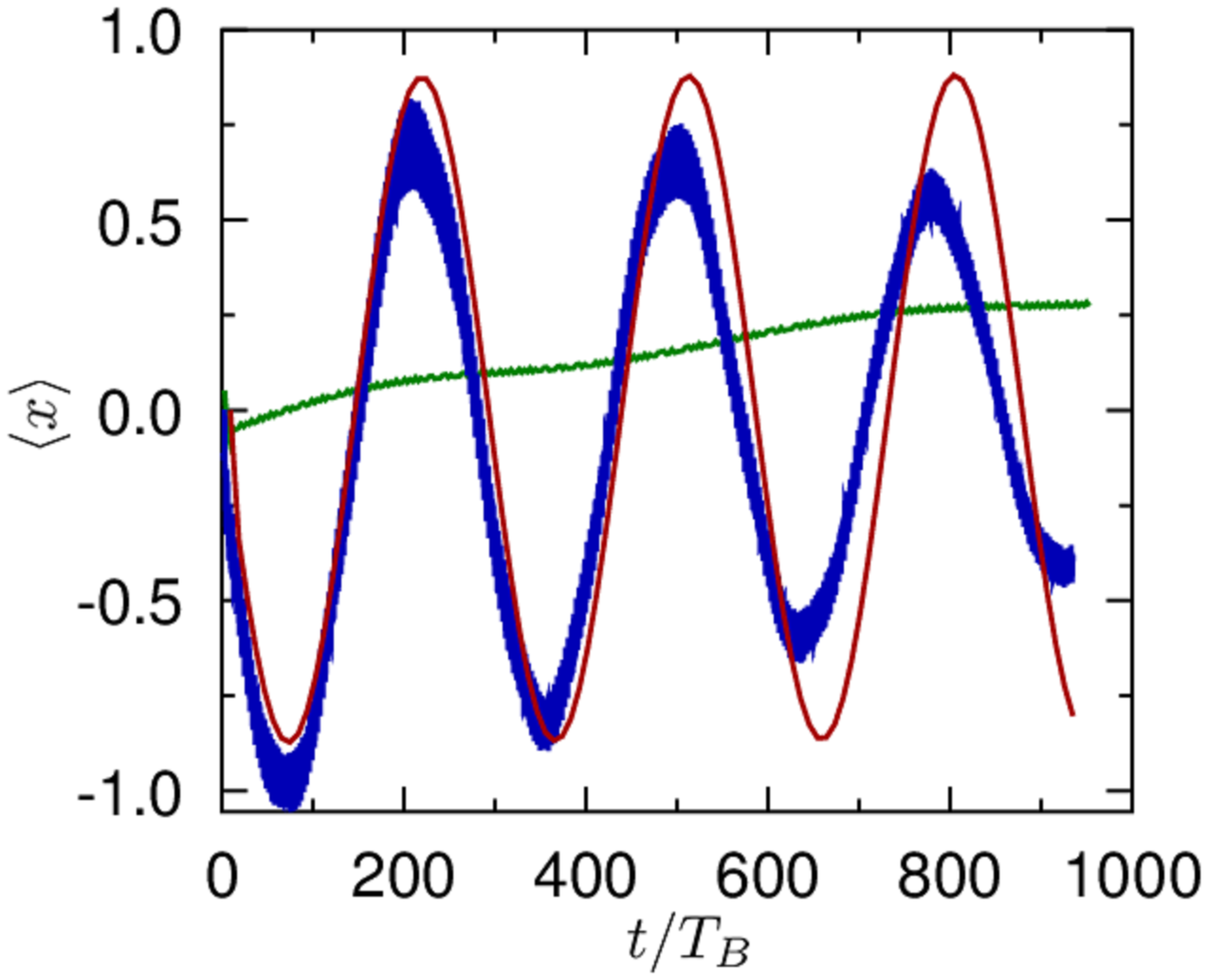} 
\par\end{centering}
\caption{\label{fig:ZB_4spinor}Evolution of the average position $\left\langle x(t)\right\rangle $
for $0\leq t\leq3T_{ZB}$. Exact Schr\"odinger equation (thick blue
line) and discrete model Eq.~(\ref{eq:Spinor4cndn}) (red line) for
an initial spinor $2^{-1}(1,1,1,1)$, $\sigma^{2}=500$ and $k_{0}=0$.
The non-oscillating green line is the exact Schr\"odinger equation
result for an initial spinor $2^{-1}(1,-1,1,1)$, which does not show
Zitterbewegung. Potential parameters are $V_{1}=6$, $F=1$, $A_{1,1}^{(1)}=-5.0\times10^{-3}i$
and $A_{1,-1}^{(1)}=-7.5\times10^{-3}i$ giving $\Omega_{1}=-1.5\times10^{-3}$.
The effective mass $E_{0}=1.6\times10^{-3}$ is generated by the potential
$V_{S}(x)=5\times10^{-3}\cos(4\pi x)$. The Zitterbewegung period
and amplitude agree with the theoretical values $T_{ZB}=2\pi/(2E_{0})=310T_{B}$
and $\left|\Omega_{1}\right|/E_{0}=0.93$. }
\end{figure}

The Zitterbewegung is described in the same way as for the spinor-2
case: 
\begin{align*}
\frac{d\left\langle x\right\rangle }{dt}= & -2\Omega_{1}\left\langle \alpha_{x}\right\rangle \\
= & -2\Omega_{1}\intop dx\left[c_{+}^{*}(x,t)d_{-}(x,t)+\mathrm{c.c}\right.\\
 & +\left.c_{-}^{*}(x,t)d(x,t)+\mathrm{c.c.}\right].
\end{align*}
In the simple case $p_{x}=0$ with a spatially broad initial wave
packet $\boldsymbol{\psi}=$$2^{-1}(a_{+},a_{-},b_{+},b-)G^{(k_{0})}(x)$
one obtains 
\[
\frac{d\left\langle x\right\rangle }{dt}=-2\Omega_{1}\left[\left(a_{+}^{*}b_{-}+a_{-}^{*}b_{+}\right)\,e^{2iE_{0}t}+\mathrm{c.c.}\right]
\]
showing an oscillation amplitude proportional to $\Omega_{1}/E_{0}$,
controlled by the initial coherence. The superposition of a ``spin
up particle'' $(a_{\text{+}},a_{-})$ and a ``spin down antiparticle''
$(b_{\text{+}},b_{-})$ $\boldsymbol{\psi}=2^{-1}(1,1,1,1)$~\footnote{This distinction is meaningful only if $p\ll mc$.},
leads to $\left\langle x(t)\right\rangle =-\left(\Omega_{1}/E_{0}\right)\sin(2E_{0}t)$.
States with $\left(a_{+}^{*}b_{-}+a_{-}^{*}b_{+}\right)=0$, for instance
$\boldsymbol{\psi}=2^{-1}(1,-1,1,1)$, display no Zitterbewegung.
These results are illustrated in Fig.~\ref{fig:ZB_4spinor}. The
oscillations (blue line) obtained from the Schr\"odinger equation
are in good agreement with the simulation of Eq.~(\ref{eq:Spinor4cndn})
displayed in red. On the time scale of a few Zitterbewegung periods
$T_{ZB}=1/2E_{0}$, diffusion effects are here negligible (one finds
$DT_{ZB}=4\Omega_{1}^{2}\pi/(E_{0}^{2}\sigma^{2})\sim10^{-2}$), but
in contrast to the spinor-2 model displayed in Fig.~\ref{fig:ZB_discret},
the exact Schr\"odinger equation shows parasitic Landau-Zener tunneling
into the continuum (due to the presence of populated excited states
$d_{\pm}$), leading to a slow decrease in the spinor negative-energy
amplitudes and thus to the oscillation amplitude. Fast, small-amplitude
Rabi oscillations at frequency $\Omega_{1}\gg T_{ZB}^{-1}$ between
ground and excited states are responsible for the apparent thickening
of the blue line in Fig.~\ref{fig:ZB_4spinor}: it is due to the
asymmetry of the excited state $\varphi_{n}^{e}(x)$ with respect
to the center of its well $n$ leading thus an average position which
differs by a fraction of a lattice step as compared to the ground
state average position. Note finally that the second initial condition
spinor $\boldsymbol{\psi}=2^{-1}(1,-1,1,1)$ (green line) do not display
Zitterbewegung, as expected.

\subsection{Spinor-4 Weyl representation }

A Dirac equation in the Weyl representation can be obtained with a
different coupling scheme. The calculation follows the same lines
as in the previous section, and we shall simply indicate the main
steps below. We use the Hamiltonian $\bar{H}$ of Eq.~(\ref{eq:Hoverbar})
with $V_{S}(x)=0$ and with the modulations
\begin{align*}
f_{2}(t) & =A_{1,1}^{(2)}e^{i\omega_{B}t}e^{i\Delta t}+A_{1,-1}^{(2)}e^{i\omega_{B}t}e^{-i\Delta t}+\mathrm{c.c.}\\
f_{1}(t) & =A_{0,1}^{(1)}e^{i\Delta t}+\mathrm{c.c}.
\end{align*}
The general developments of Sec.~\ref{sec:GeneralModel} then lead
to: 
\begin{align*}
i\frac{d}{dt}c_{n} & =T_{n,1}^{ge}d_{n+1}+T_{n,-1}^{ge}d_{n-1}+T_{n,0}^{ge}d_{n}\\
i\frac{d}{dt}d_{n} & =T_{n,1}^{eg}c_{n+1}+T_{n,-1}^{eg}c_{n-1}+T_{n,0}^{eg}c_{n}.
\end{align*}
We then choose $A_{0,1}^{(1)}$ real and define the real parameter
\begin{align*}
E_{W} & =T_{n,0}^{ge}=T_{n,0}^{eg}\\
 & =-A_{0,1}^{(1)}V_{1}\left\langle \varphi_{0}^{g}\left|\cos(2\pi x)\right|\varphi_{0}^{e}\right\rangle .
\end{align*}
Making the amplitudes $A_{1,\pm1}^{(2)}$ imaginary and tuning them
in such a way that $A_{1,1}^{(2)}\left\langle \varphi_{0}^{g}\right|\cos(\pi x)\left|\varphi_{1}^{e}\right\rangle $$=A_{-1,1}^{(2)}\left\langle \varphi_{1}^{g}\right|\cos(\pi x)\left|\varphi_{0}^{e}\right\rangle $,
one has 
\begin{align*}
T_{n,1}^{ge} & =(-1)^{n}V_{2}A_{1,1}^{(2)}\left\langle \varphi_{0}^{g}\right|\cos(\pi x)\left|\varphi_{1}^{e}\right\rangle =i(-1)^{n}\Omega_{W}\\
T_{n,-1}^{ge} & =(-1)^{n}V_{2}A_{-1,1}^{(2)}\left\langle \varphi_{0}^{g}\right|\cos(\pi x)\left|\varphi_{-1}^{e}\right\rangle =-i(-1)^{n}\Omega_{W}\\
T_{n,1}^{eg} & =T_{n,1}^{ge}\\
T_{n,-1}^{eg} & =-T_{n,1}^{ge}
\end{align*}
and thus

\begin{eqnarray}
i\frac{d}{dt}c_{n}= & (-1)^{n} & i\Omega_{W}\left(d_{n+1}-d_{n-1}\right)+E_{W}d_{n}\nonumber \\
i\frac{d}{dt}d_{n}= & -(-1)^{n} & i\Omega_{W}\left(c_{n+1}-c_{n-1}\right)+E_{W}c_{n}.\label{eq:Spinor4cndn-1}
\end{eqnarray}
The continuous limit of these two equations gives
\begin{align*}
i\partial_{t}c_{+}(x,t) & =-2\Omega_{W}p_{x}d_{-}+E_{W}d_{+}\\
i\partial_{t}d_{-}(x,t) & =-2\Omega_{W}p_{x}c_{+}+E_{W}c_{-}\\
i\partial_{t}d_{+}(x,t) & =2\Omega_{W}p_{x}c_{-}+E_{W}c_{+}\\
i\partial_{t}c_{-}(x,t) & =2\Omega_{W}p_{x}d_{+}+E_{W}d_{-}.
\end{align*}
The Weyl spinor-4 is thus defined as $\boldsymbol{\psi}_{W}(x,t)=\left(c_{+},d_{-},d_{+},c_{-}\right)$
and follows the equation $i\partial_{t}\boldsymbol{\psi}_{W}=H_{W}\boldsymbol{\psi}_{W}$
with 
\[
H_{W}=-2\Omega_{W}\left(\begin{array}{cc}
\sigma_{x}p_{x} & 0\\
0 & -\sigma_{x}p_{x}
\end{array}\right)+E_{W}\left(\begin{array}{cc}
0 & 1\\
1 & 0
\end{array}\right)
\]
which is the Dirac Hamiltonian in the Weyl representation, Eq.~(\ref{eq:H_Weyl}),
with $\boldsymbol{p}$ parallel to the $x$ axis.

\section{Prospects for an experimental realization\label{sec:ExperimentalRelization}}

The present proposal of a quantum simulator of Dirac physics depends
on techniques that are widely used experimentally. It is based on
driving of ultracold atoms by modulations of a 1D optical lattice~\cite{Cohen-TannoudjiDGO:AdvancesInAtomicPhysics::11,Eckardt:AtomicQuantumGasesPeriodicallyDrivenOptLatt:RMP17},
a technique that has been used from the early days of optical lattice
physics, from the seminal experiments of observation of Bloch oscillations~\cite{BenDahan:BlochOsc:PRL96}
and the Wannier-Stark ladder~\cite{Niu:LandauZennerWS:PRL96}, dynamical
localization and Anderson physics~\cite{Moore:LDynFirst:PRL94,Garreau:QuantumSimulationOfDisordered:CRP17},
Landau-Zener tunneling~\cite{Zenesini:LandauZener:PRL09,Creffield:ExpansionOfMatterWavesDrivenOptLat:PRA10},
to, more recently, the generation of artificial gauge fields~\cite{Struck:TunableGaugePotentialDrivenLattices:PRL12,Hauke:NonAbelianGaugeFieldsShakenOL:PRL12}.
This makes our system particularly simple, both conceptually and experimentally,
not involving, for example, Raman transitions or Zeeman-level manipulation.
The main limitation of driven systems is the loss of atoms to the
continuum via dynamic Landau-Zener coupling, which requires a careful
optimization of the parameters. However, most effects described here
survive to moderate losses, e.g. the Zitterbewegung, as it can be
seen from Figs.~\ref{fig:ZBFalseColors} and~\ref{fig:ZB_4spinor}.

Several techniques have also been developed for atom detection, recently
attaining single-site resolution thanks to the \emph{quantum gas microscope}~\cite{Greif:SiteResolvedImagingQuantumGas:S16,Haller:SingleAtomImagingOfFermionsQuantumGasMicroscope:NP17}
or near-field imaging~\cite{Zimmermann:HighResolutionImagingUltracoldAtoms:NJP11}.
For the particular situation studied here, a possible difficulty is
the necessity of distinguishing the contribution of atoms located
in even and odd sites. This can obviously be done site by site if
single-site resolution is attained. Another, potentially more practical,
way to do so is to \emph{select} atoms from even/odd sites \emph{before}
detection. A possible strategy is the following: After the desired
dynamics is studied (e.g. Zitterbewegung) the tilt of the potential
is adiabatically tuned to zero, leaving only a flat trapping potential
$\mathsf{V}_{a}\cos(2\mathsf{k}_{L}\mathit{\mathsf{x}})\exp(-\mathsf{y}^{2}/\mathsf{w}_{a}^{2})$,
where we take into account the tranverse Gaussian profile of the laser
beam. One then turns on adiabatically a transversely-shifted double-period
potential $\mathsf{V}_{b}\cos(\mathsf{k}_{L}\mathit{x}+\varphi)\exp\left[-(\mathsf{y}-\mathsf{y}_{0})^{2}/\mathsf{w}_{b}^{2}\right]$;
for $\varphi=0$ (resp.~$\pi$) this potential will mostly affect
even (resp. odd) sites. By adjusting the ratio $\mathsf{V}_{b}/\mathsf{V}_{a}$
and the shift $\mathsf{y}_{0}$ one can create a transverse ``gutter''
that induces losses in even (resp.~odd) sites. One can then either
detect the lost atoms, that is even- (resp.~odd-)site population,
or remaining atoms, i.e. odd- (resp.~even-) site population. If the
potential allows two Wannier-Stark ladders, one can adjust $\mathsf{V}_{a}$
before turning $\mathsf{V}_{b}$ on so as to induce losses in the
excited WS ladder.

As a concrete example, consider the 4-spinor $\boldsymbol{\psi}=(c_{+},c_{-},d_{+},d_{-})$
Eq.~(\ref{eq:spinor-4}). In the particle-antiparticle context, the
first component $c_{+}$ (for example) corresponds to the spin-up
component for a particle at rest. In our quantum simulator it corresponds
to the slowly varying envelope of the population of the ground ladder
odd sites. Such quantity can be measured by first lowering the potential
barrier (or increasing the slope) so that the atoms in the excited
ladder escape, and then measuring the population $\left|c_{+}(x)\right|^{2}$
using the techniques described above. For the excited ladder components
as $\left|d_{+}(x)\right|^{2}$ (odd sites), one can first remove
even-site atoms using the method presented above, then lower the lattice
depth allowing the excited-ladder atoms to escape while ground-ladder
atoms remain trapped, and one detects the atoms that are leaking.
The other components can be detected in a similar way.

\section{\label{sec:Conclusion}Conclusion}

The present work introduces a general scheme based on the Wannier-Stark
Hamiltonian, realizable with ultracold atoms in 1D optical lattices,
allowing for the quantum simulation of Dirac physics, with a great
flexibility in the choice of the parameters and of the properties
of the resulting quantum simulator. One can control the effective
mass, realize spinor-2 and spinor-4 Dirac equations both in the standard
and in the Weyl representation. Our general model opens a large field
of other possibilities which will be developed in forthcoming papers.
For instance, the spinor-4 obtained as two degenerate spinor-2 systems
can be studied in the case where the degeneracy is lifted, leading
to flat bands or to spin $3/2$-like relativistic particles. The possibilities
are even more exciting if one generalizes the above approach to higher
dimensions. In dimension 2, one can use lattice temporal modulations
to generate non-trivial artificial gauge fields~\citep{Dalibard:ArtificialGaugePotentials:RMP11,Lamata:RelativisticQuantumMechanicsTrappedIons:NJP11},
and quantum simulate the Dirac particle interaction with electromagnetic
fields (e.g. simulate the ``gyromagnetic factor'' of our ``artificial
electron''). If one uses interacting bosonic atoms in the mean-field
limit described by the Gross-Pitaevskii equation, we can study Dirac
physics in the presence of a nonlinearity, which can lead to quasiclassical
``relativistic'' chaos~\citep{Thommen:ChaosBEC:PRL03}. All these
possibilities put into evidence the power of ultracold atoms and optical
potentials as quantum simulator for a rich variety of physical systems. 
\begin{acknowledgments}
This work is supported by Agence Nationale de la Recherche (Grant
K-BEC No. ANR-13-BS04-0001-01), the Labex CEMPI (Grant No. ANR-11-LABX-0007-01),
as well as by the Ministry of Higher Education and Research, Hauts
de France council and European Regional Development Fund (ERDF) through
the Contrat de Projets Etat-Region (CPER Photonics for Society, P4S). 
\end{acknowledgments}

\appendix

\section{\label{sec:Derivation}Detailed derivation of the Dirac Hamiltonian}

This Appendix presents in more detail the calculation leading to the
coupled equations of Eqs.~(\ref{eq:GeneralEvolutionEqs}), and show
how a a Dirac-like Hamiltonian can be obtained. 

We consider here the wave packet of Eq.~(\ref{eq:Psi-general}) and
project the Schrodinger equation, $id\Psi/dt=(H_{0}+\bar{H})\Psi$
on the WS states {[}noting that $c_{n}=\left\langle \varphi_{n}^{g}\right.\left|\Psi\right\rangle \exp\left(i\omega_{B}t\right)$
and $d_{n}=\left\langle \varphi_{n}^{e}\right.\left|\Psi\right\rangle \exp\left(i\omega_{B}t+i\Delta t\right)${]}:
\begin{align}
i\frac{d}{dt}c_{n}= & \sum_{r\in\mathbb{Z}}\left\{ \left\langle \varphi_{n}^{g}\right|\bar{H}\left|\varphi_{n+r}^{g}\right\rangle e^{-ir\omega_{B}t}c_{n+r}\right.\nonumber \\
 & \left.+\left\langle \varphi_{n}^{g}\right|\bar{H}\left|\varphi_{n+r}^{e}\right\rangle e^{-ir\omega_{B}t}e^{-i\Delta t}d_{n+r}\right\} \nonumber \\
i\frac{d}{dt}d_{n}= & \sum_{r\in\mathbb{Z}}\left\{ \left\langle \varphi_{n}^{e}\right|\bar{H}\left|\varphi_{n+r}^{g}\right\rangle e^{-ir\omega_{B}t}c_{n+r}\right.\nonumber \\
 & \left.+\left\langle \varphi_{n}^{e}\right|\bar{H}\left|\varphi_{n+r}^{e}\right\rangle e^{-ir\omega_{B}t}e^{-i\Delta t}d_{n+r}\right\} \label{eq:app_evolution}
\end{align}
where the ``free evolution'' due to $H_{0}$ is canceled out. In
the following, we take as an example the particular perturbation
\begin{equation}
\bar{H}(t)=-V_{1}\cos(2\pi x)f_{1}(t)\label{eq:app_Hbar}
\end{equation}
with
\begin{equation}
f_{1}(t)=A_{1,1}^{(1)}e^{i\omega_{B}t}e^{i\Delta t}+A_{1,-1}^{(1)}e^{i\omega_{B}t}e^{-i\Delta t}+\mathrm{c.c}.\label{eq:app_modulation}
\end{equation}
The results for any other choice of Hamiltonian can be obtained along
the same lines.

From Eqs.~(\ref{eq:app_evolution}), we then have:
\begin{align}
i\frac{d}{dt}c_{n}=-V_{1} & \sum_{r\in\mathbb{Z}}\left\{ \left\langle \varphi_{n}^{g}\right|\cos2\pi x\left|\varphi_{n+r}^{g}\right\rangle f_{1}(t)e^{-ir\omega_{B}t}c_{n+r}\right.\nonumber \\
 & \left.+\left\langle \varphi_{n}^{g}\right|\cos2\pi x\left|\varphi_{n+r}^{e}\right\rangle f_{1}(t)e^{-ir\omega_{B}t}e^{-i\Delta t}d_{n+r}\right\} \nonumber \\
i\frac{d}{dt}d_{n}=-V_{1} & \sum_{r\in\mathbb{Z}}\left\{ \left\langle \varphi_{n}^{e}\right|\cos2\pi x\left|\varphi_{n+r}^{g}\right\rangle f_{1}(t)e^{-ir\omega_{B}t}e^{i\Delta t}c_{n+r}\right.\nonumber \\
 & \left.+\left\langle \varphi_{n}^{e}\right|\cos2\pi x\left|\varphi_{n+r}^{e}\right\rangle f_{1}(t)e^{-ir\omega_{B}t}d_{n+r}\right\} .\label{eq:GeneralEv}
\end{align}
We now introduce two simplifying assumptions: (\emph{i}) The overlap
integrals between WS states rapidly shrink to zero for $\left|r\right|>1$
and we can thus consider only nearest neighbor couplings, and (\emph{ii})
we neglect fast oscillations and keep only resonant contributions
in Eq.~(\ref{eq:GeneralEv}), which eliminates intra-ladder couplings
(assuming that $\Delta$ is far from $\omega_{B}$). We obtain:
\begin{align}
i\frac{d}{dt}c_{n}= & -V_{1}A_{1,1}^{(1)}\left\langle \varphi_{0}^{g}\right|\cos2\pi x\left|\varphi_{1}^{e}\right\rangle d_{n+1}\nonumber \\
 & -V_{1}A_{1,-1}^{(1)*}\left\langle \varphi_{0}^{g}\right|\cos2\pi x\left|\varphi_{-1}^{e}\right\rangle d_{n-1}\nonumber \\
i\frac{d}{dt}d_{n}= & -V_{1}A_{1,-1}^{(1)}\left\langle \varphi_{0}^{e}\right|\cos2\pi x\left|\varphi_{1}^{g}\right\rangle c_{n+1}\nonumber \\
 & -V_{1}A_{1,1}^{(1)*}\left\langle \varphi_{0}^{e}\right|\cos2\pi x\left|\varphi_{-1}^{g}\right\rangle c_{n-1},\label{eq:app_coupled_discrete}
\end{align}
that is, Eq.~(\ref{eq:general_discrete_model}) with intra-ladder
couplings off and the inter-ladder couplings of Eq.(\ref{eq:intraLadderCoupls}).
In Eq.~(\ref{eq:app_coupled_discrete}), we took into account the
reality condition of $f_{1}(t)$, $A_{1,-1}^{(1)*}=A_{-1,1}^{(1)}$,
$A_{1,1}^{(1)*}=A_{-1,-1}^{(1)}$, and the properties of overlap integrals:
\begin{align*}
\left\langle \varphi_{n}^{g}\right|\cos2\pi x\left|\varphi_{n\pm1}^{e}\right\rangle  & =\int\varphi_{n}^{g}(x)\varphi_{n\pm1}^{e}(x)\cos(2\pi x)dx\\
 & =\int\varphi_{0}^{g}(x-n)\varphi_{\pm1}^{e}(x-n)\cos(2\pi x)dx\\
 & =\left\langle \varphi_{0}^{g}\right|\cos2\pi x\left|\varphi_{\pm1}^{e}\right\rangle 
\end{align*}
and
\begin{align*}
\left\langle \varphi_{n}^{e}\right|\cos2\pi x\left|\varphi_{n\pm1}^{g}\right\rangle  & =\left\langle \varphi_{0}^{e}\right|\cos2\pi x\left|\varphi_{\pm1}^{g}\right\rangle \\
 & =\left\langle \varphi_{0}^{g}\right|\cos2\pi x\left|\varphi_{\mp1}^{e}\right\rangle 
\end{align*}
where the translational invariance of WS states was used. 

In the general framework of Sec.~\ref{sec:GeneralModel}, other contributions
to the coupling coefficients may have to be considered in Eqs.~(\ref{eq:interLadderCoupls})
and (\ref{eq:intraLadderCoupls}), and can be obtained in the same
way. Note that if a perturbation component proportional to $\cos(\pi x)$
is present, the overlap integrals are
\begin{align*}
\left\langle \varphi_{n}^{g}\right|\cos\pi x\left|\varphi_{n\pm1}^{e}\right\rangle  & =\int\varphi_{n}^{g}(x)\varphi_{n\pm1}^{e}(x)\cos(\pi x)dx\\
 & =\int\varphi_{0}^{g}(x)\varphi_{\pm1}^{e}(x)\cos(\pi x+\pi n)dx\\
 & =(-1)^{n}\left\langle \varphi_{0}^{g}\right|\cos\pi x\left|\varphi_{\pm1}^{e}\right\rangle \\
\left\langle \varphi_{n}^{e}\right|\cos\pi x\left|\varphi_{n\pm1}^{g}\right\rangle  & =\pm(-1)^{n}\left\langle \varphi_{0}^{e}\right|\cos\pi x\left|\varphi_{\pm1}^{g}\right\rangle ,
\end{align*}
and thus depend on the even or odd character of the site label $n$.

A Dirac-like equation can be derived from Eq.~(\ref{eq:app_coupled_discrete}).
If we tune the modulation coefficients such that 
\[
A_{1,1}^{(1)}\left\langle \varphi_{0}^{g}\right|\cos2\pi x\left|\varphi_{1}^{e}\right\rangle =-A_{1,-1}^{(1)*}\left\langle \varphi_{0}^{g}\right|\cos2\pi x\left|\varphi_{-1}^{e}\right\rangle 
\]
we find
\begin{align*}
i\frac{d}{dt}c_{n}= & -V_{1}A_{1,1}^{(1)}\left\langle \varphi_{0}^{g}\right|\cos2\pi x\left|\varphi_{1}^{e}\right\rangle \left[d_{n+1}-d_{n-1}\right],\\
i\frac{d}{dt}d_{n}= & A_{1,1}^{(1)*}V_{1}\left\langle \varphi_{1}^{e}\right|\cos2\pi x\left|\varphi_{0}^{g}\right\rangle \left[c_{n+1}-c_{n-1}\right]
\end{align*}
and assuming imaginary amplitudes (i.e choosing the phase of the modulations
suitably) gives
\begin{align}
i\frac{d}{dt}c_{n}= & i\Omega\left[d_{n+1}-d_{n-1}\right]\nonumber \\
i\frac{d}{dt}d_{n}= & i\Omega\left[c_{n+1}-c_{n-1}\right]\label{eq:app_Dirac_discrete}
\end{align}
where, $V_{1}A_{1,1}^{(1)}\left\langle \varphi_{0}^{g}\right|\cos2\pi x\left|\varphi_{1}^{e}\right\rangle =-i\Omega$.
Note that these equations correspond to two independent sub-lattices,
the amplitudes $c_{n}$ for $n$ odd being coupled to $d_{n}$ for
$n$ even, and conversely. 

We thus conclude that the ``suitable'' form of the potential corresponding
to Eqs.~(\ref{eq:app_Hbar}) and (\ref{eq:app_modulation}) leading
to Eq.~(\ref{eq:app_Dirac_discrete}) is
\[
f_{1}(t)=2a_{1,1}^{(1)}\sin\left(\omega_{B}t+\Delta t\right)+2a_{1,-1}^{(1)}\sin\left(\omega_{B}t-\Delta\right)
\]
where $a_{1,\pm1}=-iA_{1,\pm1}$ are real amplitudes with relative
weight obeying $a_{1,1}^{(1)}\left\langle \varphi_{0}^{g}\right|\cos2\pi x\left|\varphi_{1}^{e}\right\rangle $$=a_{1,-1}\left\langle \varphi_{0}^{g}\right|\cos2\pi x\left|\varphi_{-1}^{e}\right\rangle $. 

We can take the continuous limit of these equations assuming that
the amplitudes $c_{n},$ $d_{n}$ are slowly varying at the scale
of the lattice step. We can then introduce the smooth envelopes associated
to each sub-lattice: $c_{n}(t)=c_{\pm}(x=n,t)$ (the sign $\pm$ corresponding
to $n$ odd or even) and $d_{n}(t)=d_{\pm}(x=n,t)$ . We then get
\begin{align*}
i\partial_{t}c_{\pm}= & i2\Omega\frac{\partial d_{\mp}(x,t)}{\partial x}\\
i\partial_{t}d_{\pm}= & i2\Omega\frac{\partial c_{\mp}(x,t)}{\partial x}
\end{align*}
This last expression written as a Dirac equation for a massless particle
corresponding to Eq.~(\ref{eq:Dirac-4}) with $E_{0}=0$.


%

\end{document}